\documentclass[twocolumn]{jpsj3} 
\usepackage{graphicx}
\usepackage{bm}
\usepackage{color}

\title{
Unconventional Multi-gap Superconductivity and Antiferromagnetic Spin Fluctuations  
in New Iron-arsenide LaFe$_{2}$As$_{2}$  in Heavily Electron-doped Regime 
}

\author{
Takayoshi Kouchi$^{1}$, Mitsuharu Yashima$^{1}$, Hidekazu Mukuda$^{1}$\thanks{E-mail: mukuda@mp.es.osaka-u.ac.jp},\\ Shigeyuki Ishida$^{2}$, Hiroshi Eisaki$^{2}$, Yoshiyuki Yoshida$^{2}$, Kenji Kawashima$^{3}$, and Akira Iyo$^{2}$
}

\inst{
$^{1}$Graduate School of Engineering Science, Osaka University, Toyonaka, Osaka 560-8531\\
$^{2}$National Institute of Advanced Industrial Science and Technology (AIST), Tsukuba, Ibaraki 305-8568\\
$^{3}$IMURA Material R \& D Co., Ltd., Kariya, Aichi 448-0032\\
}

\abst{
We report $^{75}$As-NMR/NQR results on new  iron-arsenide compounds (La$_{0.5-x}$Na$_{0.5+x}$)Fe$_{2}$As$_{2}$.
The parent compound $x$=0 exhibits a stripe-type antiferromagnetic (AFM) order below $T_{\rm N}$=130 K.
The measurement of nuclear spin relaxation rate at hole-doped $x$=+0.3 and heavily electron-doped $x$=$-$0.5  revealed that the normal-state properties are dominated by AFM spin fluctuations (AFMSFs), which are more significant at  $x$=+0.3 than at  $x$=$-$0.5. 
Their superconducting (SC) phases are characterized by unconventional multi-gap SC state, where the smaller SC gaps are particularly weaken in common.  
The experimental results indicate the close relationship between the AFMSFs and the SC from the hole-doped state to heavily electron-doped state, 
which shed  light on a unique  SC phase emerged in the heavily electron-doped regime  being formally equivalent to non-SC compound Ba(Fe$_{0.5}$Co$_{0.5}$)$_{2}$As$_{2}$.
}

\recdate{\today}
\newpage
\begin{document}
\maketitle



After the discovery of superconductivity (SC) in the iron (Fe)-pnictide LaFeAsO$_{1-x}$F$_x$ ($T_c$=26 K)\cite{Kamihara2008},  higher SC states have been reported in Fe-based compounds in the heavily electron-doped regime, such as monolayer FeSe ($T_c\sim$65 K)\cite{Wang2012,He},  intercalated FeSe systems ($T_c$= 30$\sim$50 K)\cite{Guo,Zhao2015}, and hydrogen-substituted LaFeAsO$_{1-x}$H$_{x}$ ($T_c\sim$36 K)\cite{Iimura2012}, and so on. 
These electronic structures are dominated by large electron Fermi surface (FS), which differs from that of typical parent compounds composed of hole and electron FSs in similar sizes\cite{Mazin,Kuroki}. 
Toward the universal understanding of SC mechanism, it is important to unveil the essential features of various Fe-based compounds in extensive doping regions.
Recently, a new iron-arsenide superconductor  LaFe$_{2}$As$_{2}$ ($T_c\sim$12.1 K) in heavily electron-doped regime was discovered in  new 122-series  (La$_{0.5-x}$Na$_{0.5+x}$)Fe$_{2}$As$_{2}$ \cite{J.Q.Yang,Iyo2018,Iyo2019}. 
It is surprising because the formal valence of Fe is expected to be +1.5, being equivalent to  Ba(Fe$_{0.5}$Co$_{0.5}$)$_{2}$As$_{2}$ where no SC  was reported\cite{Sefat2009,Liu}. 
Its doping level is also equivalent to that of  reemergent antiferromagnetic (AFM) order phase of LaFeAs(O$_{0.5}$H$_{0.5}$) in 1111-series\cite{Hiraishi2014}. 
It is expected that the substitution of La$^{3+}$ with Na$^{+}$ in block layer of  (La$_{0.5-x}$Na$_{0.5+x}$)Fe$_{2}$As$_{2}$ enables us to control the doping level of conductive FeAs layers\cite{J.Q.Yang,Iyo2018}. 
In fact, the formal valence of  Fe at $x$=0 is expected to be  +2 in average, which coincides with that of typical parent Fe-pnictides. 
The compounds for $x>$0.15  shows the SC phase with a maximum $T_c$(=27 K) at $x$=+0.3\cite{Iyo2018}, that would be in the hole-doping regime.
Therefore, a new series of Fe-pnictides (La$_{0.5-x}$Na$_{0.5+x}$)Fe$_{2}$As$_{2}$ gives a unique opportunity to study the key elements for SC  in heavily electron-doped regime, and the electron-hole symmetry in the Fe-based SCs continuously by controlling a single parameter ($x$), i.e., the substitution at block layer.

In this Letter, we report $^{75}$As-NMR/NQR results on (La$_{0.5-x}$Na$_{0.5+x}$)Fe$_{2}$As$_{2}$.
The normal-state properties are dominated by AFM spin fluctuations (AFMSFs), which are more significant at hole-doped $x$=+0.3 than at heavily electron-doped $x$=$-$0.5. 
The SC phases at $x$=+0.3 and $-$0.5 are characterized by unconventional SC state with multiple gaps where the smaller SC gap is particularly weaken  in common.  
These experimental findings can be consistently accounted for by recent band calculation\cite{Usui2019,Mazin2019}, which suggests the remaining of hole FS  as a result of the mixing of the La and Fe orbitals even in the heavily electron-doped state. 
These results  indicate the close relationship between the AFMSFs and the SC state from the hole-doped to heavily electron-doped states.  



NMR/NQR measurements were performed on coarse-powder polycrystalline samples of (La$_{0.5-x}$Na$_{0.5+x}$) Fe$_{2}$As$_{2}$ with nominal contents at $x$=$-$0.5, 0, and +0.3, which were synthesized using a cubic-anvil-type high pressure apparatus\cite{Iyo2018,Iyo2019}. 
The bulk $T_c$s were determined to be $T_c\sim$9.4 K for $x$=$-$0.5 and $T_c\sim$27 K for $x$=+0.3 by SC diamagnetism in  {\it ac}-susceptibility  using an {\it in-situ} NQR coil. 
Here the $T_c$ of $x$=$-$0.5 was slightly lower than the value reported by onset of zero-resistivity  and $dc$-susceptibility in the literature\cite{Iyo2019}.  
The parent compound ($x$=0) exhibits no SC transition and an anomaly in resistivity  at $T_s$=130 K.
Nuclear spin-lattice relaxation rate  $1/T_1$  for  $x$=0 and +0.3 was measured by $^{75}$As-NMR at $B_0\sim$8 T, which was determined by fitting a recovery curve for $^{75}$As nuclear magnetization  to a multiple  exponential function $m(t)=0.1\exp(-t/T_1) + 0.9\exp(-6t/T_1)$. 
The $1/T_1$  for $x$=$-$0.5 was measured by  $^{75}$As-NQR in zero external field, 
in which the $1/T_1$ was determined by fitting a recovery curve  to a single exponential function 
$m(t)= \exp (- 3t/T_1)$.


\begin{figure}[tbp]
\centering
\includegraphics[width=7cm]{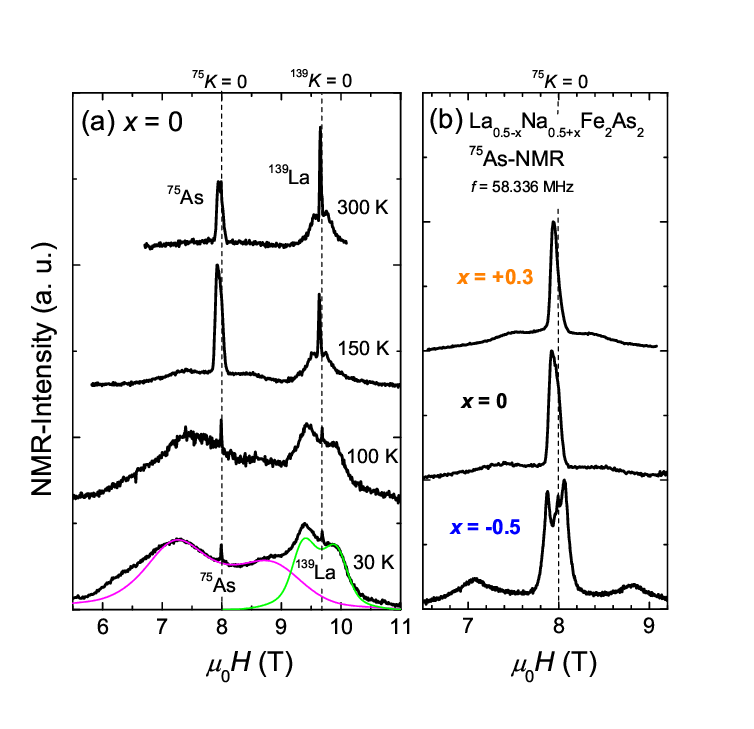}
\includegraphics[width=7cm]{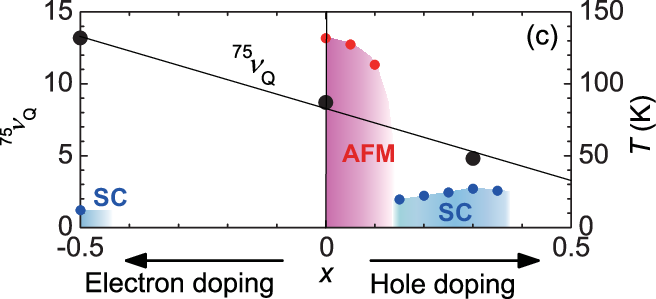}
\caption[]{(Color online)  
(a) $T$ dependence of NMR spectra for $x$=0, which exhibits broadening below $T_{\rm N}$=130 K. 
Solid curves at 30 K show $^{75}$As and $^{139}$La-NMR spectra  simulated by assuming $^{75}B_{\rm int}\sim$1.6 T and $^{75}\nu_{Q}$  at  As site, and $^{139}B_{\rm int}\sim$0.5 T and $^{139}\nu_{Q}$ at  La site. 
(b) $^{75}$As-NMR spectra for $x$=$-$0.5, 0, and +0.3 in the paramagnetic states. 
(c) The value of  $^{75}\nu_Q$ estimated from analyses of (b)  varies linearly with $x$, indicating that the formal valence of FeAs layer is controlled linearly by a single parameter  $x$ from hole-doping  to  electron-doping region. 
}
\label{spectrum}
\end{figure}

\begin{figure}[tbp]
\centering
\includegraphics[width=7cm]{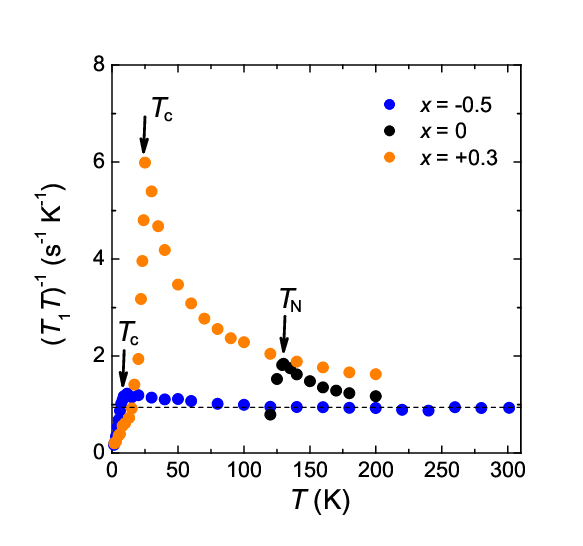}
\caption[]{(Color online)  
$T$  dependence of $1/T_1T$ probed by $^{75}$As-NMR/NQR  for $x$= +0.3, 0, $-$0.5.
The $1/T_1T$ for $x$=0 shows a peak at   $T_{\rm N}$=130 K. 
The $1/T_1T$ for $x$=+0.3 increases significantly upon cooling due to the presence of strong AFMSFs in the normal state. 
It is noteworthy that  such AFMSFs  remain even at  $x$=$-$0.5 in heavily electron-doped regime. 
}
\label{T1-N}
\end{figure}

Figure \ref{spectrum}(a) shows  $^{75}$As- ($I$=3/2)  and $^{139}$La- ($I$=7/2) NMR spectra for  $x$=0  obtained by sweeping the magnetic field ($B$) at a fixed frequency 58.336 MHz. 
The spectra at high temperatures, which are articulated due to the nuclear quadrupole interaction, exhibits a significant broadening below 130 K, indicating the appearance of uniform internal field $B_{\rm int}$ at the As and La sites in association with the AFM order at Fe site.
The $^{75}$As-NMR spectrum at 30 K can be simulated by assuming $^{75}B_{\rm int}\sim$1.6 T and the   NQR frequency $^{75}\nu_{Q}\sim$8.7 MHz. 
Noted that $^{75}B_{\rm int}$ is  comparable to those of typical parent Fe-pnictides in the stripe AFM order, such as BaFe$_{2}$As$_{2}$(Ba122) \cite{Kitagawa2008} and LeFeAsO\cite{MukudaNQR,MukudaFe2}. 
Assuming the hyperfine coupling constant at As site $^{75}A_{\rm hf}\sim$1.72 T/$\mu_{\rm B}$ reported in Ba122\cite{Kitagawa2008}, the magnetic moment ($M_{\rm AFM}$) of the Fe site   is deduced to be $\sim$0.9 $\mu_{\rm B}$.
The N\'eel temperature ($T_{\rm N}$) is evaluated to be $T_{\rm N}$=130 K from a sharp peak of  $1/T_1T$, as shown in Fig. \ref{T1-N}, which is also similar to that of Ba122. 
As for La site, $^{139}$La-NMR spectrum well below $T_{\rm N}$ can be simulated by assuming $^{139}B_{\rm int}\sim$0.5 T and $^{139}\nu_{Q}\sim$1.5 MHz, as indicated in Fig.  \ref{spectrum}(a). 
The $^{139}B_{\rm int}$ at La site  is twice larger than 0.25 T for LaFeAsO\cite{MukudaNQR}. 


Figure \ref{spectrum}(b) shows the $x$ dependence of  $^{75}$As-NMR spectra in the paramagnetic states. 
The values of $^{75}\nu_{Q}$  evaluated from the spectral analyses are plotted as a function of $x$ in Fig. \ref{spectrum}(c).
Since the $^{75}\nu_{\rm Q}$  is in proportion to an electric field gradient derived from the charge distribution around the $^{75}$As nucleus, the linear relation between  $^{75}\nu_{\rm Q}$ and $x$ is attributed to the monotonous variation of the lattice parameters and the averaged valence of the blocking layer composed of  (La$^{3+}_{0.5-x}$Na$^{+}_{0.5+x}$)\cite{Iyo2019}.  
This gives rise to a microscopic evidence that formal valence of FeAs layer can be  controlled by tuning $x$ from heavily electron-doped to hole-doped states continuously while keeping the  uncollapsed tetragonal structure. 

\begin{figure}[tbp]
\centering
\includegraphics[width=7cm]{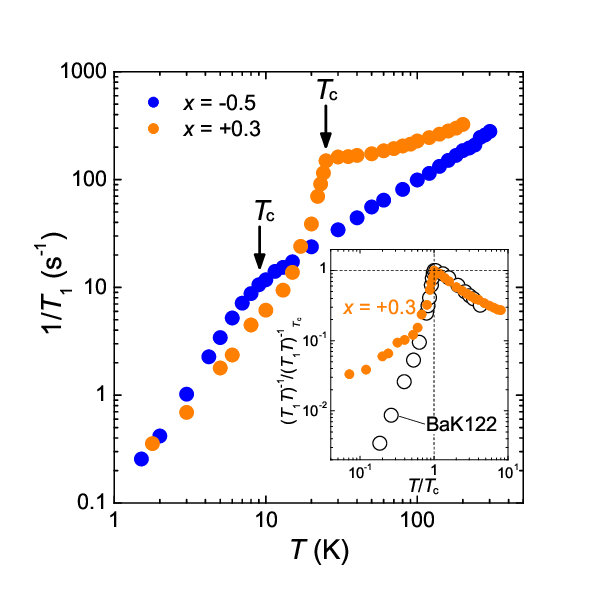}
\caption[]{(Color online)  
$T$ dependence of $1/T_1$ for $x$=$-$0.5 and +0.3, which decreases steeply  just below $T_c$  without  a coherence peak and shows a change of the slope in its $T$ dependence well below $T_c$(also see Fig. \ref{T1-2}(a,b)), which are the indications of the unconventional SC state with multiple SC gaps in both compounds. 
Inset shows the plot of  $1/T_1T$ normalized by the values at $T_{c}$ as a function of $T/T_{c}$ for $x$=+0.3 in comparison with typical hole-doped compound BaK122 ($T_c$=38 K)\cite{Yashima2009}.
}
\label{T1-SC}
\end{figure}

Figure \ref{T1-N} shows the $T$  dependence of $1/T_1T$  for $x$= +0.3, 0, $-$0.5. 
In general, $1/T_1T$ is proportional to $\sum_{\bm q} |A_{\bm q}|^2 \chi''({\bm q},\omega_0)/\omega_0$, where $A_{\bm q}$ is a wave-vector (${\bm q}$)-dependent hyperfine-coupling constant, $\chi({\bm q},\omega)$ a dynamical spin susceptibility, and $\omega_0$ an NMR/NQR frequency.  
At $x$=+0.3, the $1/T_1T$ increases significantly upon cooling toward its $T_c$, which resembles  that of typical hole-doped  compound Ba$_{0.6}$K$_{0.4}$Fe$_{2}$As$_{2}$(BaK122) with $T_c$=38 K\cite{Yashima2009,Fukazawa}, as compared in the inset of Fig.  \ref{T1-SC}.
It indicates that the normal-state property in these hole-doped states is dominated by the strong AFMSFs with finite wave vector  at around ${\bf Q}$=($\pi$, 0) and (0, $\pi$) at low energies, as widely seen in  optimally hole-doped Fe-pnictides\cite{Yashima2009,Fukazawa,Hirano,Cui}. 
Note that the enhancement of AFMSFs originates from the interband nesting of  hole and electron FSs\cite{Mazin,Kuroki} and their electron correlation effects enhanced when pnictogen height ($h_{pn}$) from Fe-plane becomes high\cite{Miyake_U,Misawa}.

\begin{figure}[htbp]
\centering
\includegraphics[width=7cm]{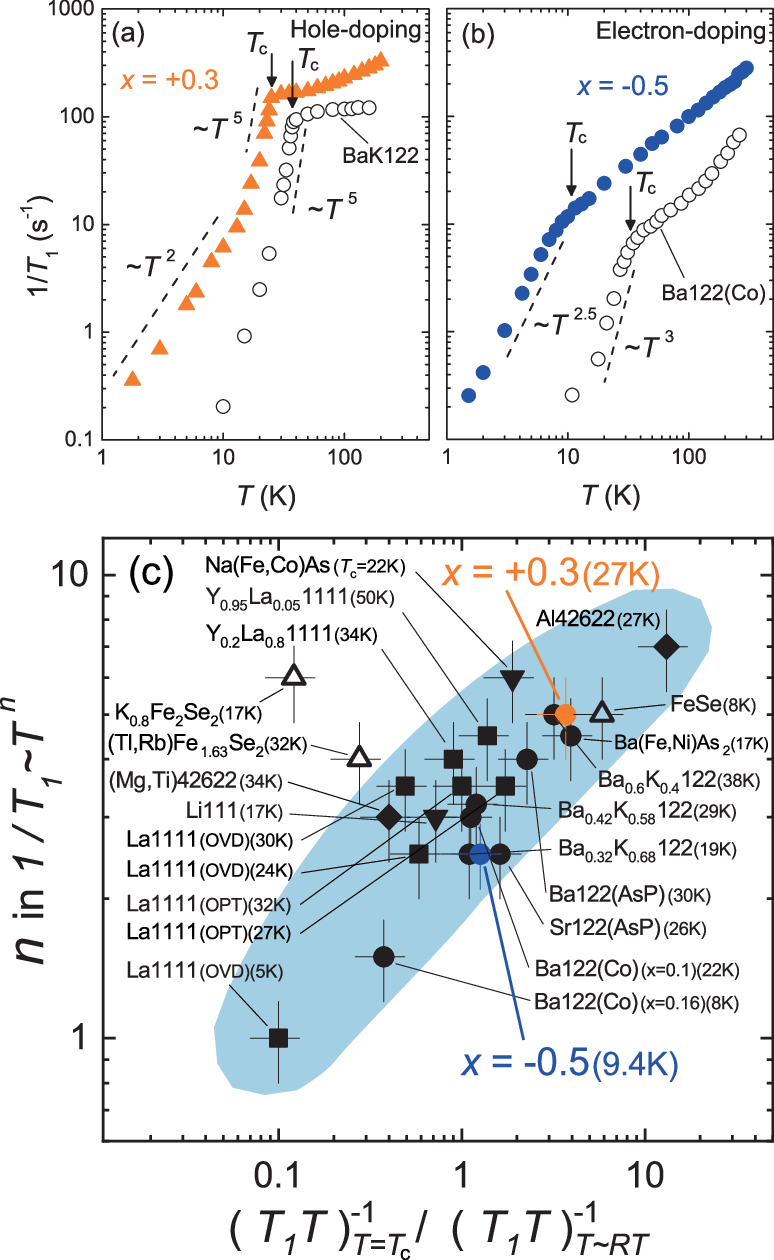}
\caption[]{(Color online)
$T$ dependence of $1/T_1$ for (a)  $x$=$+$0.3 and (b)  $x$=$-$0.5, compared with  hole-doped  BaK122 ($T_c$=38 K)\cite{Yashima2009} and  electron-doped  Ba122(Co) (22 K)\cite{NingBa122Co}, respectively. 
(c) The $n$ in the formula of  $1/T_1 \sim T^n$ (SC state) is plotted against  $(T_1T)^{-1}_{T_c}/(T_1T)^{-1}_{RT}$ (normal state). 
The data are cited from the previous report\cite{MukudaPRL} and the other literatures for Na(Fe,Co)As ($T_c$=22 K)\cite{MaPRB}, K$_{x}$Fe$_2$Se$_2$(17 K)\cite{KotegawaKFe2Se2}, (Tl,Rb)Fe$_{2}$Se$_2$ (32 K)\cite{TlRbFeSe}, La1111(OVD) (24 K, 30 K)\cite{Yang2015}, La1111(OPT) (27 K)\cite{Oka2012}, (Mg,Ti)42622 (34 K)\cite{Yamamoto2012}, Ba(Fe,Ni)$_2$As$_2$ (17 K)\cite{Chou2013}, and FeSe (8 K)\cite{Baek,FeSe}.
Here, $(T_1T)^{-1}_{T_{\rm c}}/(T_1T)^{-1}_{RT}>$1 indicates how large the AFMSFs develop from RT to $T_c$.
The $n$ in the SC state is determined in the $T$ range of 0.5$T_{\rm c} < T < T_{\rm c}$. 
The plot means that as AFMSFs are more significantly enhanced, the reduction rate in $1/T_1$ just below $T_c$ is more remarkable, suggesting the close relationship between the AFMSFs  and the onset of sign-reversing SC.
}
\label{T1-2}
\end{figure}

In the SC state of $x$=+0.3, the $1/T_1$ decreases markedly without any trace of a Hebel-Slichter coherence peak just below $T_c$  and shows a change of the slope in its $T$ dependence below $\sim0.5 T_c$,  as shown in Fig. \ref{T1-SC}, which are the indications of the unconventional SC state with sign-reversing multiple SC gaps composed of relatively larger and smaller gaps\cite{Ding}. 
Assuming the power $n$ in $1/T_1\propto T^n$ for the SC state, the $n$ is close to $\sim$5 just below $T_c$  analogous to that in BaK122\cite{Yashima2009}, as compared in Fig. \ref{T1-2}(a). 
The feature just below $T_c$ is dominated primarily by the larger SC gap within  the multiple gaps, indicating  the similarity in the larger  gap structure for $x$=+0.3 and BaK122.  
By contrast, the $n$ decreases to $\sim$2 approximately below $\sim$0.5$T_c$, in contrast with the case of BaK122 characterized by the nodeless SC gaps in all the FSs\cite{Ding,Yashima2009}.
It suggests that the smaller SC gap of $x$=+0.3 is significantly weaken or suppressed, namely, the quasiparticles relating to a smaller gap are predominantly affected by pair breaking effect due to some impurities or possible disorders. 
This unfavorable situation for SC may be one of the reasons that the $T_c$ in $x$=+0.3 is  lower than that in BaK122.

Next we address the results of the $1/T_1T$ at $x$=$-$0.5 probed by $^{75}$As-NQR at $f$=13.5 MHz without external field. 
As shown in Fig. \ref{T1-N},  the $1/T_1T$ at $x$=$-$0.5 is slightly enhanced in the normal state upon cooling toward the $T_{c}$, suggesting that the AFMSFs at low energies are weak but present even in the heavily electron-doped regime.
It is quite surprising because the formal valence of Fe is expected to be +1.5, which is equivalent to the doping level  of  Ba(Fe$_{0.5}$Co$_{0.5}$)$_{2}$As$_{2}$, where no AFMSFs and no SC were reported\cite{Sefat2009,Liu,Ning}. 
It is in contrast with the cases of many electron-doped Fe-pnictides that the AFMSFs at low energies are particularly suppressed\cite{MukudaFe2,Ning,KotegawaKFe2Se2}, suggesting that the {\it effective} electron-doping level of  $x$=$-$0.5 may be different from the value expected from formal valence.
In the SC state, the $1/T_1$ decreases below $T_{c}\sim9.4$ K, as shown in Fig. \ref{T1-2}(b).
The $1/T_1$ does not show the coherence peak just below $T_c$, and follows the $1/T_1\propto T^n$ with approximately $n\sim$2.5 below $T_c$, which gradually changes to $n\sim$2 at further low temperatures. 
These features can be also understood by the framework of  unconventional SC state with multiple gaps, which is characterized by much broader gap structures for most of the SC gaps, as deduced from the previous analyses in the literature\cite{Yashima2009}. 

In many Fe-pnictide SCs, there are close relationship between strong (weak) AFMSFs in the normal state and the steep (gradual) decrease in $1/T_1$ in the SC state\cite{MukudaPRL}. 
Figure \ref{T1-2}(c) shows   the $n$ in  $1/T_1 \sim T^n$ in the SC state plotted as a function of  $(T_1T)^{-1}_{T_{\rm c}}/(T_1T)^{-1}_{RT}$ in the normal state for various Fe-pnictide SCs\cite{MukudaPRL}.  
Here, $(T_1T)^{-1}_{T_{\rm c}}$ and $(T_1T)^{-1}_{RT}$ are the values at $T$=$T_{\rm c}$ and room temperature(RT)\cite{FeSe}, respectively, and hence the $(T_1T)^{-1}_{T_{\rm c}}/(T_1T)^{-1}_{RT}>$1 indicates how large the AFMSFs develop from RT to $T_c$.
Here $n$ is evaluated in the $T$ range of 0.5 $T_{\rm c} < T < T_{\rm c}$ that represents the suppression of the coherence factor and the size in the larger SC gaps within the multiple gaps leading to the SC transition primarily\cite{MukudaPRL}.
The data of $x$=+0.3 and $-$0.5 are similar to that of typical hole-doped BaK122 and electron-doped Ba(Fe$_{1-x}$Co$_{x}$)$_{2}$As$_{2}$(=Ba122(Co)) in Figs. \ref{T1-2}(a) and \ref{T1-2}(b), respectively, which are on the empirical  tendency for many Fe-based SCs as shown in Fig. \ref{T1-2}(c) \cite{MukudaPRL}.  
This plot points to the experimental  fact that as AFMSFs become more dominant in the normal state, the reduction rate in $1/T_1$ just below $T_{\rm c}$, i.e. $n$, increases more significantly, suggesting the close relationship between the AFMSFs and sign-reversing SC states among many Fe-pnictide SCs.
We note that the data of heavily electron-doped $A_{x}$Fe$_2$Se$_2$($A$=K,Rb,Tl) without hole FSs\cite{Y.Zhang2011,T.Qian2011}  deviate from this tendency, whereas the datum of FeSe with hole and electron FSs is plotted on it\cite{FeSe}.   
Thus we suggest that the empirical relation of Fig. \ref{T1-2} holds in many sign-reversing Fe-based SCs with hole and electron FSs even though their FS sizes are not exactly the same.  
The deviation from this relation in $A_{x}$Fe$_2$Se$_2$ originates from the lack of low energy AFMSFs in  the NMR measurement\cite{KotegawaKFe2Se2,TorchettiKFe2Se2,TlRbFeSe}, implying that it is necessary to study the spin fluctuations at higher energies using neutron scattering measurements for further universal understanding of the wide variety of Fe-based SCs.

Finally, we address a possible reason for the onset of SC in heavily electron-doped LaFe$_2$As$_2$.
Recently band calculation on $x$=$-$0.5 (LaFe$_2$As$_2$) in uncollapsed tetragonal phase suggested that the hybridization of La-$5d$ and Fe-$3d$ orbitals gives rise to the remaining of a hole FS derived from $d_{xy}$ orbital around the $\Gamma$ point in spite of such heavily electron-doped state\cite{Usui2019,Mazin2019}.
It gives a possible interpretation for the presence of weak AFMSFs and unconventional SC phase at $x$=$-$0.5 in the experiment. 
According to this scenario, the other experimental fact that the smaller gap within multiple SC gaps are weaken both in $x$=+0.3 and $-$0.5 may be attributed to the effect of accidental mixing of the La-$5d$ orbitals into Fe-$3d$ orbitals that is unfavorable for SC. 
In this context, the remaining of a hole FS from $d_{xy}$ orbital reminds us of the reemergent SC phase ($x$=0.3$\sim$0.4) and AFM order phase ($x$=0.5) of LaFeAs(O$_{1-x}$H$_{x}$) in heavily electron-doped regime\cite{Iimura2012,Hiraishi2014}, which derives from the fact that the $h_{pn}$ was kept at high position\cite{Suzuki_H}. 
Although the reason for the remaining of the hole FS is different,  it is worth noting that the small hole FS at the Fermi level ($E_{\rm F}$) with nearly cylindrical feature is one of the key elements for the onset of unconventional SC in heavily electron-doped region. 
The non-SC state in Ba(Fe$_{0.5}$Co$_{0.5}$)$_{2}$As$_{2}$ can be ascribed to the lack of the features such as no cylindrical FS, and no electron correlations of Fe-$3d$ electrons\cite{Sefat2009,Ning}, where the $h_{pn}$ becomes low monotonously and the rigid band picture may be applicable as increasing the doping level\cite{Liu}.
The possible presence of hole FS in LaFe$_2$As$_2$ would cause the real {\it effective} doping level to be different from the formal one.



In summary,  we performed $^{75}$As-NMR/NQR studies on heavily electron-doped ($x$=$-$0.5), parent ($x$=0), and hole doped ($x$=+0.3) states of  (La$_{0.5-x}$Na$_{0.5+x}$)Fe$_2$As$_2$. 
The linear relation between $^{75}$As-NQR frequency and $x$ ensures microscopically that the valence of blocking layer is continuously controlled from $x$=$-$0.5 to +0.3 while keeping the uncollapsed tetragonal structure.   
The $x$=0 exhibits  a stripe-type AFM order below $T_{\rm N}$=130 K, which is comparable to BaFe$_2$As$_2$.   
The hole-doped state at $x$=+0.3 is dominated by strong AFMSFs in the normal state, and characterized by the unconventional SC state with multiple gaps, as reported in (Ba$_{0.6}$K$_{0.4}$)Fe$_2$As$_2$, whereas the smaller gap of $x$=+0.3 is significantly weaken. 
In heavily electron-doped state of $x$=$-$0.5, we observed  weak AFMSFs in the normal state and  unconventional multiple-gapped SC state characterized by obscure gap structures.
Recent band calculation suggested the presence of hole FS  derived from $d_{xy}$ orbital as a result of the mixing of the La-$5d$ and Fe-$3d$ orbitals even in the heavily electron-doped state\cite{Usui2019,Mazin2019}. 
It would  give rise to accidental gap-minima that may be a possible origin of the smaller gaps affected by the disorders in this experiment.
The experiment at further low temperatures is needed to determine whether the gap-nodes are present  or not in the future, since the  $1/T_1\propto T$ term  derived from the residual density of states at $E_{\rm F}$  in the Fe-based SCs with nodes\cite{NakaiPRL,Dulguun,Miyamoto,KinouchiPRB} was not seen even down to $T\sim1.5$ K.  
Consequently, we remark that these experimental results reveal the close relationship between the presence of AFMSFs and onset of sign-reversing SC states, which are consistently accounted for by the spin-fluctuation-based SC mechanism.



{\footnotesize 
We thank H. Usui and K. Kuroki for valuable discussion. 
This work was supported by JSPS KAKENHI Grant Nos. 16H04013  and 18K18734. }



\begin{thebibliography}{99} 

\bibitem{Kamihara2008} Y. Kamihara, T. Watanabe, M. Hirano, and H. Hosono, J. Am. Chem. Soc. {\bf 130}, 3296 (2008).
\bibitem{Wang2012} Q. Y. Wang, Z. Li, W. H. Zhang, Z. C. Zhang, J. S. Zhang, W. Li, H. Ding, Y. B. Ou, P. Deng, K. Chang, J. Wen, C. L. Song, K. He, J. F. Jia, S. H. Ji, Y. Y. Wang, L. L. Wang, X. Chen, X. C. Ma, and Q. K. Xue, Chin. Phys. Lett. {\bf 29}, 037402 (2012).
\bibitem{He} S. He, J. He, W. Zhang, L. Zhao, D. Liu, X. Liu, D. Mou, Y. Ou, Q. Wang, Z. Li, L. Wang, Y. Peng, Y. Liu, C. Chen, L. Yu, G. Liu, X. Dong, J. Zhang, C. Chen, Z. Xu, X. Chen, X. Ma, Q.-K. Xue, and X. J. Zhou, Nat. Mater. {\bf 12}, 605 (2013).
\bibitem{Zhao2015} L. Zhao, A. Liang, D. Yuan, Y. Hu, D. Liu, J. Huang, S. He, B. Shen, Y. Xu, X. Liu, L. Yu, G. Liu, H. Zhou, Y. Huang, X. Dong, F. Zhou, K. Liu, Z. Lu, Z. Zhao, C. Chen, Z. Xu, and X. J. Zhou, Nat. Commun. {\bf 7}, 10608 (2015).
\bibitem{Guo} J. Guo, S. Jin, G. Wang, S. Wang, K. Zhu, T. Zhou, M. He, and X.
Chen, Phys. Rev. B {\bf  82}, 180520(R) (2010).
\bibitem{Iimura2012} S. Iimura, S. Matsuishi, H. Sato, T. Hanna, Y. Muraba, S. W. Kim, J. E. Kim, M. Takata, and H. Hosono, Nat. Commun. {\bf 3}, 943 (2012).
\bibitem{Mazin} I. I. Mazin, D. J. Singh, M. D. Johannes, and M. H. Du, Phys. Rev. Lett.  {\bf 101}, 057003 (2008).
\bibitem{Kuroki} K. Kuroki, S. Onari, R. Arita, H. Usui, Y. Tanaka, H. Kontani, and H. Aoki, Phys. Rev. Lett.  {\bf 101}, 087004 (2008).
\bibitem{J.Q.Yang} J. -Q. Yang, S. Nandi, B. Saparov, P. \v{C}erm\'{a}k, Y. Xiao, Y. Su, W. T. Jin, A. Schneidewind, Th. Br\"{u}ckel, R. W. McCallum, T. A. Lograsso, B. C. Sales, and D. G Mandrus, Phys. Rev. B {\bf 91}, 024501 (2015).
\bibitem{Iyo2018} A. Iyo, K. Kawashima, S. Ishida, H. Fujiwara, Y. Gotoh, H. Eisaki, and Y. Yoshida, J. Am. Chem. Sci. {\bf 140}, 369 (2018).
\bibitem{Iyo2019} A. Iyo, S. Ishida, H. Fujiwara, Y. Gotoh, I. Hase, Y. Yoshida, H. Eisaki, and K. Kawashima, J. Phys. Chem. Lett. {\bf 10}, 1018 (2019).
\bibitem{Sefat2009} A. S. Sefat, D. J. Singh, R. Jin, M. A. McGuire, B. C. Sales, and D. Mandrus, Phys. Rev. B {\bf  79}, 024512  (2009).
\bibitem{Liu} C. Liu, A. D. Palczewski, R. S. Dhaka, T. Kondo, R. M. Fernandes, E. D. Mun, H. Hodovanets, A. N. Thaler, J. Schmalian, S. L. Bud'ko, P. C. Canfield, and A. Kaminski, Phys. Rev. B {\bf 84}, 020509(R) (2011).
\bibitem{Hiraishi2014} M. Hiraishi, S. Iimura, K. M. Kojima, J. Yamaura, H. Hiraka, K. Ikeda, P. Miao, Y. Ishikawa, S. Torii, M. Miyazaki, I. Yamauchi, A. Koda, K. Ishii, M. Yoshida, J. Mizuki, R. Kadono, R. Kumai, T. Kamiyama, T. Otomo, Y. Murakami, S. Matsuishi, and H. Hosono, Nat. Phys. {\bf 10}, 300 (2014).
\bibitem{Usui2019} H. Usui  and K. Kuroki,  arXiv:1905.05488.
\bibitem{Mazin2019} I. I. Mazin, M. Shimizu, N. Takemori, and H. O. Jeschke, arXiv:1905.06190.


\bibitem{Kitagawa2008} K. Kitagawa, N. Katayama, K. Ohgushi, M. Yoshida, and M. Takigawa, J. Phys. Soc. Jpn. {\bf 77}, 114709 (2008).
\bibitem{MukudaNQR} H. Mukuda, N. Terasaki, H. Kinouchi, M. Yashima, Y. Kitaoka, S. Suzuki, S. Miyasaka, S. Tajima, K. Miyazawa, P. M. Shirage, H. Kito, H. Eisaki, and A. Iyo, J. Phys. Soc. Jpn. {\bf 77}, 093704 (2008).
\bibitem{MukudaFe2} H. Mukuda, N. Terasaki, N. Tamura, H. Kinouchi, M. Yashima, Y. Kitaoka, K. Miyazawa, P. M. Shirage, S. Suzuki, S. Miyasaka, S. Tajima, H. Kito, H. Eisaki, and A. Iyo, J. Phys. Soc. Jpn. {\bf 78}, 084717 (2009).

\bibitem{Yashima2009} M. Yashima, H. Nishimura, H. Mukuda, Y. Kitaoka, K. Miyazawa, P. M. Shirage, K. Kihou, H. Kito, H. Eisaki, and A. Iyo, J. Phys. Soc. Jpn. {\bf 78}, 103702 (2009).
\bibitem{Fukazawa} H. Fukazawa, T. Yamazaki, K. Kondo, Y. Kohori, N. Takeshita, P. M. Shirage, K. Kihou, K. Miyazawa, H. Kito, H. Eisaki, and A. Iyo, J. Phys. Soc. Jpn.  {\bf 78}, 033704 (2009).
\bibitem{Hirano} M. Hirano, Y. Yamada, T. Saito, R. Nagashima, T. Konishi, T. Toriyama, Y. Ohta, H. Fukazawa, Y. Kohori, Y. Furukawa, K. Kihou, C.-H. Lee, A. Iyo, and H. Eisaki, J. Phys. Soc. Jpn. {\bf 81}, 054704 (2012).
\bibitem{Cui} J. Cui, Q.-P. Ding, W. R. Meier, A. E. Bohmer, T. Kong, V. Borisov, Y. Lee, S. L. Bud'ko, R. Valent\'i, P. C. Canfield, and Y. Furukawa, Phys. Rev. B {\bf 96}, 104512 (2017).
\bibitem{Miyake_U} T. Miyake, K. Nakamura, R. Arita, and M. Imada, J. Phys. Soc. Jpn.  {\bf 79}, 044705 (2010).
\bibitem{Misawa} M. Hirayama, T. Misawa, T. Miyake, and M. Imada, J. Phys. Soc. Jpn. {\bf 84}, 093703 (2015) .
\bibitem{Ding} H. Ding, P. Richard, K. Nakayama, K. Sugawara, T. Arakane, Y. Sekiba, A. Takayama, S. Souma, T. Sato, T. Takahashi, Z. Wang, X. Dai, Z. Fang, G. F. Chen, J. L. Luo, and N. L. Wang, Europhys. Lett. {\bf 83}, 47001 (2008).
\bibitem{Ning}F. L. Ning, K. Ahilan, T. Imai, A. S. Sefat, M. A. McGuire, B. C. Sales, D. Mandrus, P. Cheng, B. Shen, and H. -H. Wen, Phys. Rev. Lett. {\bf 104}, 037001 (2010).
\bibitem{KotegawaKFe2Se2} H. Kotegawa, Y. Hara, H. Nohara, H. Tou, Y. Mizuguchi, H. Takeya, and Y. Takano, J. Phys. Soc. Jpn. {\bf 80}, 043708 (2011).
\bibitem{NingBa122Co} F. Ning, K. Ahilan, T. Imai, A. S. Sefat, R. Jin, M. A. McGuire, B. C. Sales, and D. Mandrus, J. Phys. Soc. Jpn. {\bf 78}, 013711 (2009).
\bibitem{MukudaPRL} H. Mukuda, S. Furukawa, H. Kinouchi, M. Yashima, Y. Kitaoka, P. M. Shirage, H. Eisaki, and A. Iyo, Phys. Rev. Lett. {\bf 109}, 157001 (2012).
\bibitem{MaPRB} L. Ma, J. Dai, P. S. Wang, X. R. Lu, Y. Sung, C. Zhang, G. T. Tan, P. Dai, D. Hu, S. L. Li, B. Normand, and W. Yu, Phys. Rev. B {\bf 90}, 144502 (2014).
\bibitem{TlRbFeSe} L. Ma, G. F. Ji, J. Zhang, J. B. He, D. M. Wang, G. F. Chen, W. Bao, and W. Yu, Phys. Rev. B {\bf 83}, 174510 (2011).
\bibitem{Yang2015} J. Yang, R. Zhou, L. L. Wei, H. X. Yang, J. Q. Li, Z. X. Zhao, and G. Q. Zheng, Chin. Phys. Lett. {\bf 32}, 107401 (2015).
\bibitem{Oka2012} T. Oka, Z. Li, S. Kawasaki, G. F. Chen, N. L. Wang, and G. Q. Zheng, Phys. Rev. Lett. {\bf 108}, 047001 (2012).
\bibitem{Yamamoto2012} K. Yamamoto, H. Mukuda, H. Kinouchi, M. Yashima, Y. Kitaoka, M. Yogi, S. Sato, H. Ogino, and J. Shimoyama, J. Phys. Soc. Jpn. {\bf 81}, 053702 (2012).
\bibitem{Chou2013} R. Zhou, Z. Li, J. Yang, D. L. Sun, C. T. Lin, and G. Q. Zheng, Nat. Commun. {\bf 4}, 2265 (2013).
\bibitem{Baek} S. -H. Baek, D. V. Efremov, J. M. Ok, J. S. Kim, J. van den Brink, and B. B\"{u}chner, Nat. Mater. {\bf 14}, 210 (2015).
\bibitem{FeSe}  In  the case of FeSe ($T_c$=8 K),  the value of $(T_1T)^{-1}_{RT}$ was substituted by the value at 90 K, since it increases below 90 K distinctly\cite{Baek}.
\bibitem{Y.Zhang2011} Y. Zhang, L. X. Yang, M. Xu, Z. R. Ye, F. Chen, C. He, H. C. Xu, J. Jiang, B. P. Xie, J. J. Ying, X. F. Wang, X. H. Chen, J. P. Hu, M. Matsunami, S. Kimura, and D. L. Feng, Nat. Mater. {\bf 10}, 273 (2011).
\bibitem{T.Qian2011} T. Qian, X. -P. Wang, W. -C. Jin, P. Zhang, P. Richard, G. Xu, X. Dai, Z. Fang, J. -G. Guo, X. -L. Chen, and H. Ding, Phys. Rev. Lett. {\bf 106}, 187001 (2011).
\bibitem{TorchettiKFe2Se2} D. A. Torchetti, M. Fu, D. C. Christensen, K. J. Nelson, T. Imai, H. C. Lei, and C. Petrovic, Phys. Rev. B {\bf 83}, 104508 (2011).
\bibitem{Suzuki_H} K. Suzuki, H. Usui, S. Iimura, Y. Sato, S. Matsuishi, H. Hosono, and K. Kuroki, Phys. Rev. Lett. {\bf 113}, 027002 (2014).
\bibitem{NakaiPRL} Y.~Nakai, T.~Iye, S.~Kitagawa, K.~Ishida, H.~Ikeda, S.~Kasahara, H.~Shishido, T.~Shibauchi, Y.~Matsuda, and T.~Terashima, Phys. Rev. Lett. {\bf 105}, 107003 (2010).
\bibitem{Dulguun} T. Dulguun, H. Mukuda, T. Kobayashi, F. Engetsu, H. Kinouchi, M. Yashima, Y. Kitaoka, S. Miyasaka, and S. Tajima, Phys. Rev. B {\bf 85}, 144515 (2012).
\bibitem{Miyamoto} M. Miyamoto, H. Mukuda, T. Kobayashi, M. Yashima, Y. Kitaoka, S. Miyasaka, and S. Tajima, Phys. Rev. B {\bf 92}, 125154 (2015).
\bibitem{KinouchiPRB} H. Kinouchi, H. Mukuda, Y. Kitaoka, P. M. Shirage, H. Fujihisa, Y. Gotoh, H. Eisaki, and A. Iyo, Phys. Rev. B {\bf 87}, 121101(R) (2013).


\end{thebibliography}
\end{document}